%% file: CoreShellSpher_en.tex
\documentclass[12pt]{iopart}

\usepackage{graphicx}
\usepackage{iopams}

\begin{document}

\title[Tamm Plasmon in a Structure with the Nanocomposite]
{Tamm Plasmon in a Structure with the Nanocomposite Containing Spheroidal Core-Shell Particles}

\author{P S Pankin$^{1,2,3,4}$, S Y Vetrov$^{2,3}$, and I V Timofeev$^{3,2}$}

\address{$^1$Federal Research Center “Krasnoyarsk Scientific Center, Russian Academy of Sciences, Siberian Branch”, Krasnoyarsk, 660036 Russia}

\address{$^2$Siberian Federal University, Institute of Engineering Physics and Radio Electronics, Krasnoyarsk, 660041 Russia}

\address{$^3$Kirensky Institute of Physics, Federal Research Center “Krasnoyarsk Scientific Center, Russian Academy of Sciences, Siberian Branch”, Krasnoyarsk, 660036 Russia}

\address{$^4$Siberian Federal University, Polytechnic Institute, Krasnoyarsk, 660041 Russia}

\ead{p.s.pankin@mail.ru}

\begin{abstract}

Spectral peculiarities of the structure consisting of a photonic crystal coated with a nanocomposite have been investigated. The nanocomposite used contains spheroidal nanoparticles with a dielectric core and a metallic shell, which are uniformly dispersed in a transparent matrix. The spectral manifestation of the observed Tamm plasmon polariton and Fabry--Perot mode has been examined. A significant polarization sensitivity of the spectra upon variation in the nanoparticle shape has been demonstrated. The dispersion curves presented for the Tamm plasmon polariton and Fabry--Perot mode are shown to be in good agreement with the spectra obtained by the transfer matrix method.   

\end{abstract}

\noindent{\it Keywords}: Tamm plasmon, nanocomposite, photonic crystal, core-shell particles




\section{Introduction}

A Tamm plasmon polariton (TPP) is a surface mode implemented when light is blocked between two mirrors \cite{Kaliteevski2007, Vinogradov2006}, one of which is characterized by the Bragg reflection and represents a photonic crystal (PC) and the other exhibits the metal-type reflection and has a negative permittivity:  $Re(\varepsilon) < 0$. A TPP manifests itself in the form of resonance lines in the reflectance or transmittance spectra of a structure \cite{Sasin2008}. 
In contrast to a surface plasmon polariton (SPP), a TPP can be excited for both the TM and TE polarizations of light, even at its normal incidence onto the interface.
TPPs have found application in lasers \cite{Symonds2012}, single-photon sources \cite{gazzano2012}, heat emitters \cite{ChenKuoPing2017}, and in enhancement of the nonlinear effects \cite{Afinogenov2018} and Purcell effect \cite{Kaliteevski2018}. 

The TPP mode can be hybridized with the modes of other types under their simultaneous excitation in a system. Previously, the electric-field \cite{gessler2014electro} and temperature \cite{grossmann2011tuneable} control by the hybrid TPP-exciton modes \cite{Symonds2009} was proposed and a single-photon source on their basis was developed \cite{braun2015enhanced}. The hybrid TPP-SPP modes \cite{Afinogenov2013} were used to design a sensor sensitive to the refractive index of a material coating a metallic film \cite{das2014tamm} and to enhance the luminescence of molecules placed onto this film \cite{chen2014tamm}. The hybridization of the TPP and localized surface plasmon \cite{Liu2012a} makes it possible to enhance the localization of light on the metal nanowire surface. In study \cite{pankin2017tunable}, we showed the possibility of electric-field and temperature control of the hybrid TPP-microcavity modes, which were used to design novel solar cells \cite{Zhang2013}, white light-emitting diodes \cite{Zhang2015}, and metal-dielectric PCs \cite{Kaliteevski2015}.

The additional opportunities of controlling the PC spectrum, which are obtained by embedding nanocomposite (NC) layers in the PC structure were earlier studied experimentally \cite{husaini2011plasmon, husaini2012enhanced} and theoretically \cite{Avdeeva2011mt, Moiseev2012, Pankin2014mt, Moiseev2014, dadoenkova2017optical}. The NCs, including color glass as the simplest example \cite{Karmakar2016}, consist of a transparent matrix with dispersed metallic nanoparticles. The NC permittivity is described well in the effective medium approximation \cite{Sihvola1999bk, Vinogradov2008}.
In a certain wavelength range, an NC can exhibit the metal-type optical properties and serve as a mirror for forming the TPP \cite{Bikbaev2013mt}. The permittivity of an NC mirror can be tuned by changing the nanoparticle shape, volume fraction, and material, which is an undoubtable advantage of such a mirror over the metallic one. 

Previously, we discussed the possibility of forming two TPPs in two photonic band gaps (PBGs) of a PC \cite{Pankin2016m_JOpt}. We examined the spectral manifestation of the TPP in a structure containing the NC layer with spherical particles consisting of a dielectric core and a metallic shell \cite{Fang2008, GhoshChaudhuri2012}. In this study, in addition to \cite{Pankin2016m_JOpt}, we numerically and analytically investigate the variation in the spectra with the change in the particle shape from spherical to spheroidal \cite{Penninkhof2008} and demonstrate the anisotropy of the reflectance spectra and the occurrence of Fabry--Perot modes, along with the TPP, in the spectrum.   

\section{Model}

The model under study consists of a PC conjugated with the NC layer (Fig.\ref{fig:Structure}a). The PC consisting of $N = 15$ periods is formed by alternating layers of materials with refractive indices of $n_a = 2$ and $n_b = 1.5$. All the layers have a quarter-wave thickness of $n_a d_a=n_b d_b=422$ nm. In addition, the structure contains the top PC layer with a refractive index of $n_a = 2$ and a thickness of $d_{Top} = 170$ nm adjacent to the NC. The NC layer with a thickness of $d_{NC} = 130$ nm consists of nanoparticles uniformly dispersed in a transparent matrix with a refractive index of $n_m = \sqrt{\varepsilon_m} = 1.5$. Nanoparticles are formed from a dielectric core with a refractive index of $n_c = \sqrt{\varepsilon_c} = 2.65$ and a metallic shell and have the form of confocal ellipsoids of revolution (Fig. \ref{fig:Structure}b). The silver shell refractive index $n_s = \sqrt{\varepsilon_s}$ is expressed through the permittivity taken in the Drude--Sommerfeld approximation

\begin{equation}
\varepsilon _s \left( \omega \right)=\varepsilon _\infty -\frac{\omega _p^2 }{\omega \left( {\omega +i\gamma } \right)}.
\label{MG:eqDrude}
\end{equation} 
Here, $\varepsilon _\infty$ is the constant taking into account the contributions of interband transitions of bound electrons, $\omega _{p}$ is the plasma frequency, $\gamma $ is the reciprocal electron relaxation time, and $\omega$ is the frequency of incident light. For silver, we have $\varepsilon _\infty$~=~5, $\omega _{p}$~=~9 eV, and $\gamma $~=~0.02~ eV \cite{Johnson_Christy1972}. The structure is placed in air with a refractive index of $n_{Air} = 1$.

\begin{center}
\includegraphics[scale=1]{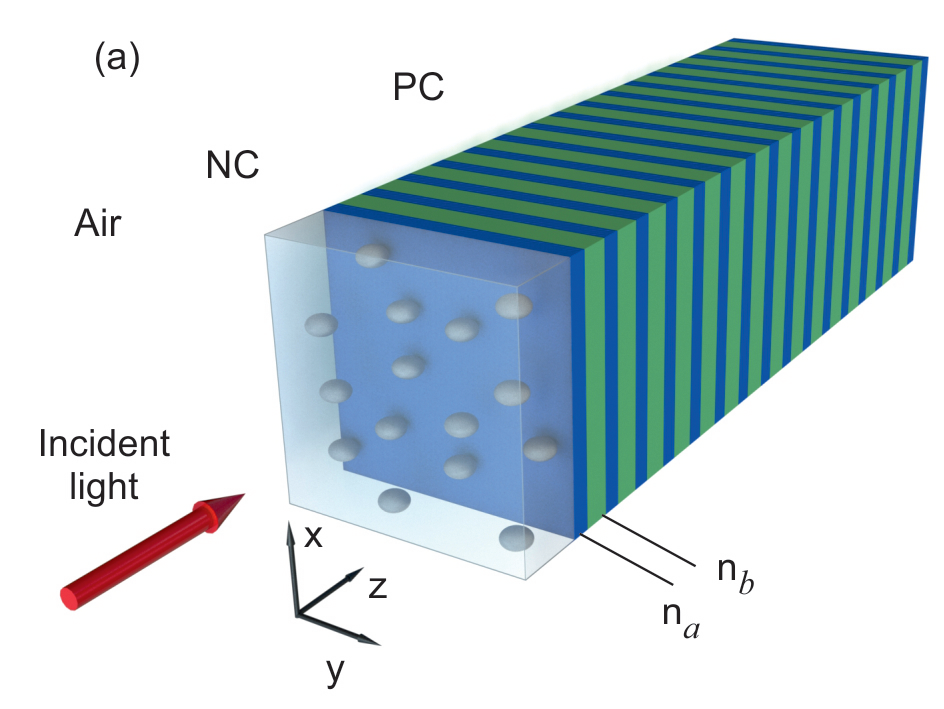}
\includegraphics[scale=1]{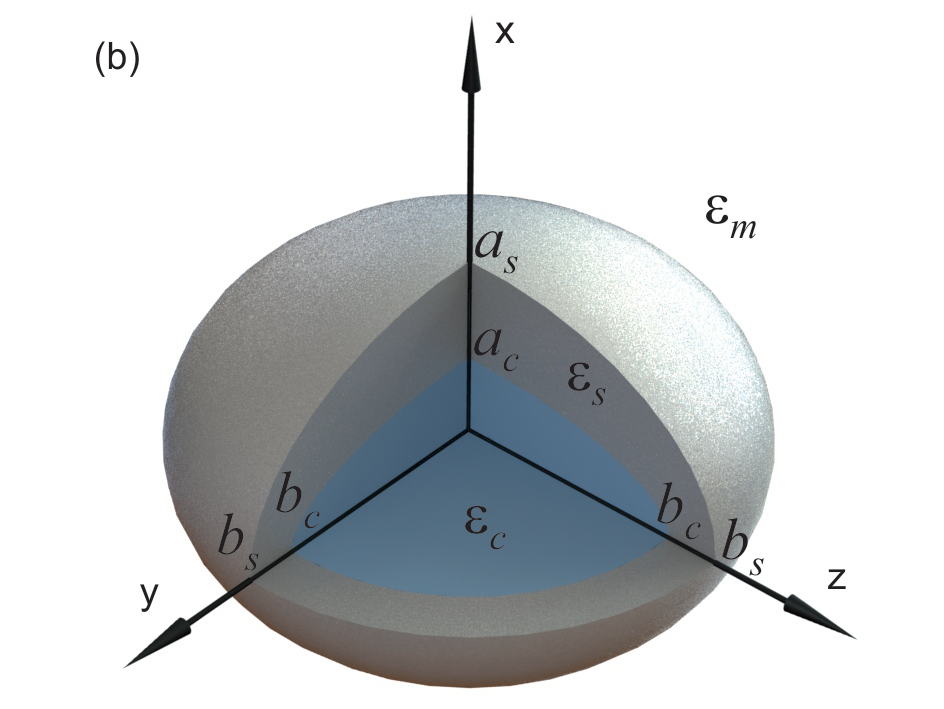}\\
Figure 1. (a) Photonic crystal conjugated with a nanocomposite. (b) Nanoparticle cross section. 
\label{fig:Structure}
\end{center}

The propagation of a light wave in an NC is calculated in the Maxwell--Garnett effective medium approximation. This approach suggests the quasistationary consideration of an NC layer with a nanoparticle size much smaller than the light wavelength in a medium. The polar axes of nanospheroids are assumed to be oriented along the $x$ axis of the system. In this case, the NC is a uniaxial media with the effective permittivity in the form of a diagonal tensor with the components $\varepsilon_{xx}$ and $\varepsilon_{yy} = \varepsilon_{zz}$ \cite{sihvola1990CoreShellEllipsoid}:

\begin{equation}
\varepsilon_{xx,yy} = \varepsilon_{m}(1 + \frac{f \alpha^{\prime}_{x,y}}{1 -L_{s}^{x,y}  f  \alpha^{\prime}_{x,y}});
\label{eq:MG}
\end{equation}
\begin{equation}
\alpha^{\prime}_{x,y} = \frac{\left(\varepsilon _s-\varepsilon _m\right)\left(\varepsilon _s+L_{c}^{x,y}(\varepsilon _c - \varepsilon _s\right))+\beta\left(\varepsilon _c-\varepsilon _s\right)\left(\varepsilon _s+L_{s}^{x,y}(\varepsilon _m - \varepsilon _s\right))}{\left(\varepsilon _m+L_{s}^{x,y}(\varepsilon _s - \varepsilon _m\right))\left(\varepsilon _s+L_{c}^{x,y}(\varepsilon _c - \varepsilon _s\right)) + \beta L_{s}^{x,y}(1-L_{s}^{x,y})\left(\varepsilon _c-\varepsilon _s\right)},
\end{equation}
where $\beta  = a_c b_c^{2}/a_s b_s^{2}$ is the ratio between the particle core volume and the entire particle volume and $f$ is the filling factor, i. e., the volume fraction of nanoparticles in the composite. The depolarization factors are expressed by the formulas
\begin{equation}
L^{x}_{c,s}=\frac{1}{1-\xi _{c,s}^2}(1-\xi_{c,s} \frac{\arcsin(\sqrt{1-\xi_{c,s}^2})}{\sqrt{1-\xi_{c,s}^2}});
\end{equation}
\begin{equation}
L^{y}_{c,s}= (1-L^{x}_{c,s})/2,
\end{equation}
where $\xi_{c,s} = a_{c,s}/b_{c,s}$  is the ratio between the lengths of the polar and equatorial axes of the core or shell.

Let us consider the behavior of the NC permittivity during flattening of a spherical particle with radius $r$ along the axis of rotation $x$. We assume the core and shell volumes to be invariable, their shape to stay spheroidal after flattening, and the confocality condition to be satisfied. Hereinafter, the quantities normalized to the initial radius are marked with a tilde: $\tilde a = a/r$. Then, with a decrease in the polar semi-axis $a_s$ of the shell, its equatorial semi-axis increases in accordance with the law $\tilde b_s = 1/\sqrt{\tilde a_s}$. The polar semi-axis $\tilde a_c = \beta/ \tilde b^{2}_{c}$ of the core is expressed through the equatorial semi-axis $b_c$, which is determined as a real positive root of the equation

\begin{equation}
\tilde b^{6}_{c} - \tilde b^{4}_{c} (\tilde b^{2}_{s}  - \tilde a^{2}_{s} ) - \beta^{2} = 0.
\end{equation}

Figure \ref{fig:Permittivity} shows the real part $\varepsilon_{xx,yy}^{\prime} = Re(\varepsilon_{xx,yy})$ of NC permittivity tensor components \ref{eq:MG} as a function of the ratio $\xi_{s}$ between the particle axes. It can be seen that at $\xi_{s} = 1$, i. e., for spherical particles, the NC represents an isotropic optical material with $\varepsilon_{xx} = \varepsilon_{yy}$. When the particle shape is different from spherical, the behavior of the permittivity tensor components $\varepsilon_{xx}$ (Fig. \ref{fig:Permittivity}a) and $\varepsilon_{yy}$ (Fig. \ref{fig:Permittivity}b) is essentially different: the NC becomes optically anisotropic. It can be seen that the NC permittivity has two resonance regions corresponding to two surface plasmons localized at both sides of the particle metallic shell \cite{Brandl2007, klimov2014nanoplasmonics}. Since we assume the shell volume to stay invariable during flattening, the latter leads to a decrease in the shell thickness. This enhances the coupling between the localized surface plasmons, which  can be observed as repulsion of the resonance regions of the NC permittivity. 

\begin{center}
\includegraphics[scale=1]{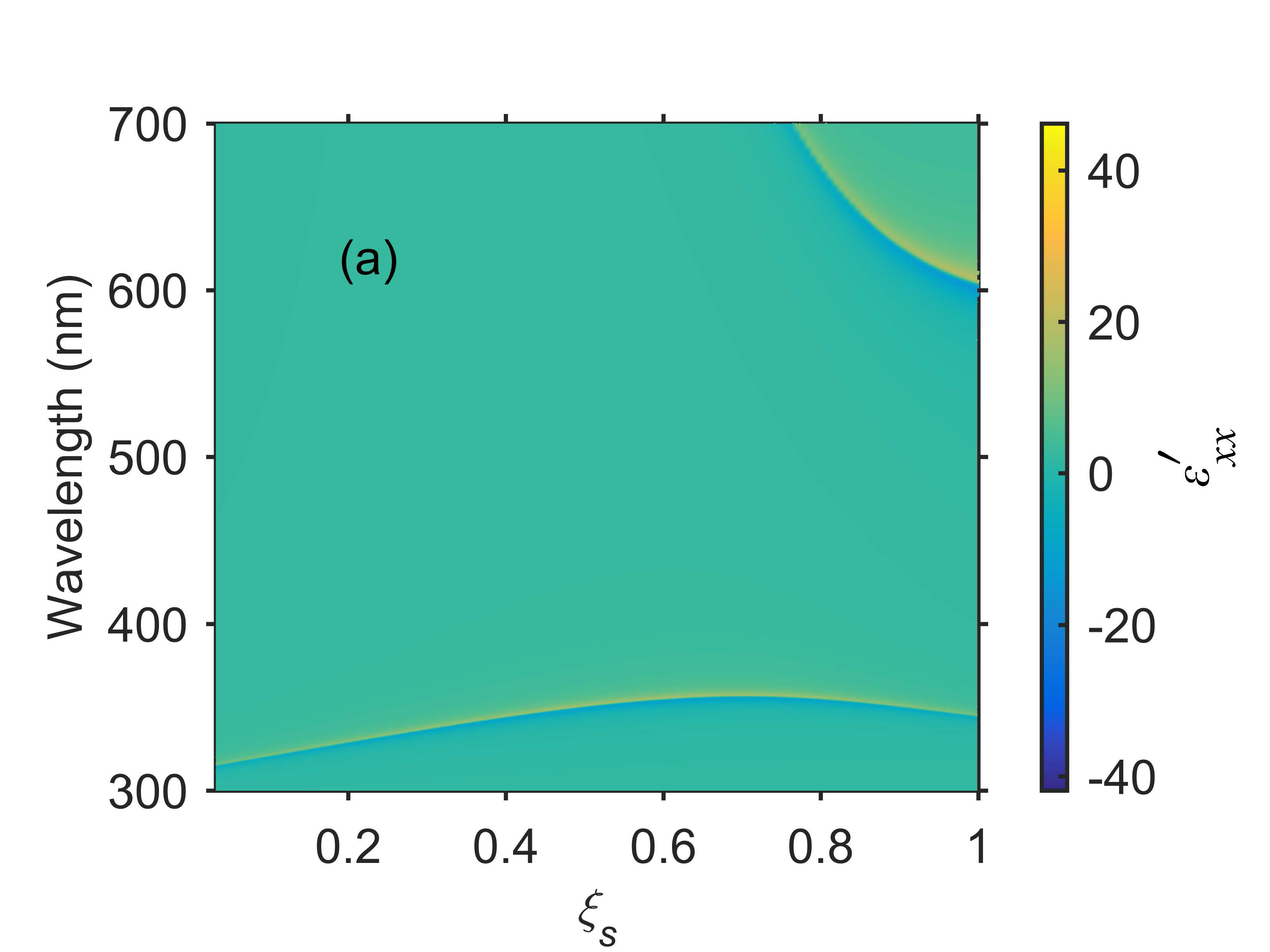}
\includegraphics[scale=1]{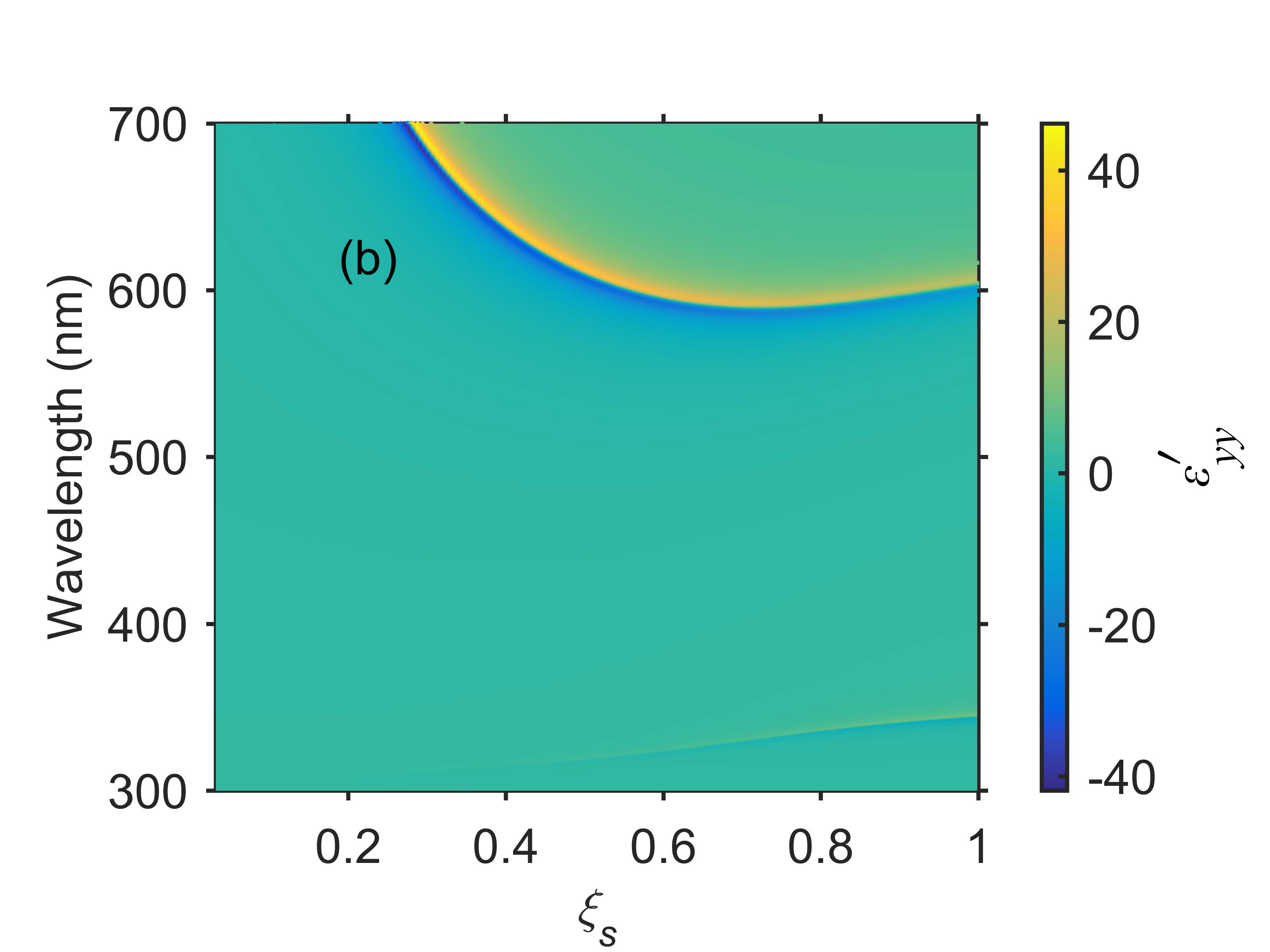}\\
Figure 2. Real part of the NC permittivity tensor components (a) $\varepsilon_{xx}$ and (b) $\varepsilon_{yy}$ calculated using formula \ref{eq:MG} as a function of the ratio $\xi_{s}$ between particle axes at $\beta = 0.275$ and $f=0.08$. 
\label{fig:Permittivity}
\end{center}

\section{Results and discussion}

We study the reflectance spectra of the structure shown in Fig. \ref{fig:Structure}a during particle flattening in dependence of the ratio $\xi_{s}$ between the particle axes. Figure \ref{fig:Spectra} presents the reflectance spectra calculated by the transfer matrix method \cite{Yeh1979, bethune1989optical} for the normally incident plane light wave polarized along the $x$ (Fig. \ref{fig:Spectra}a) or $y$ axis (Fig. \ref{fig:Spectra}b). 

\begin{center}
\includegraphics[scale=1]{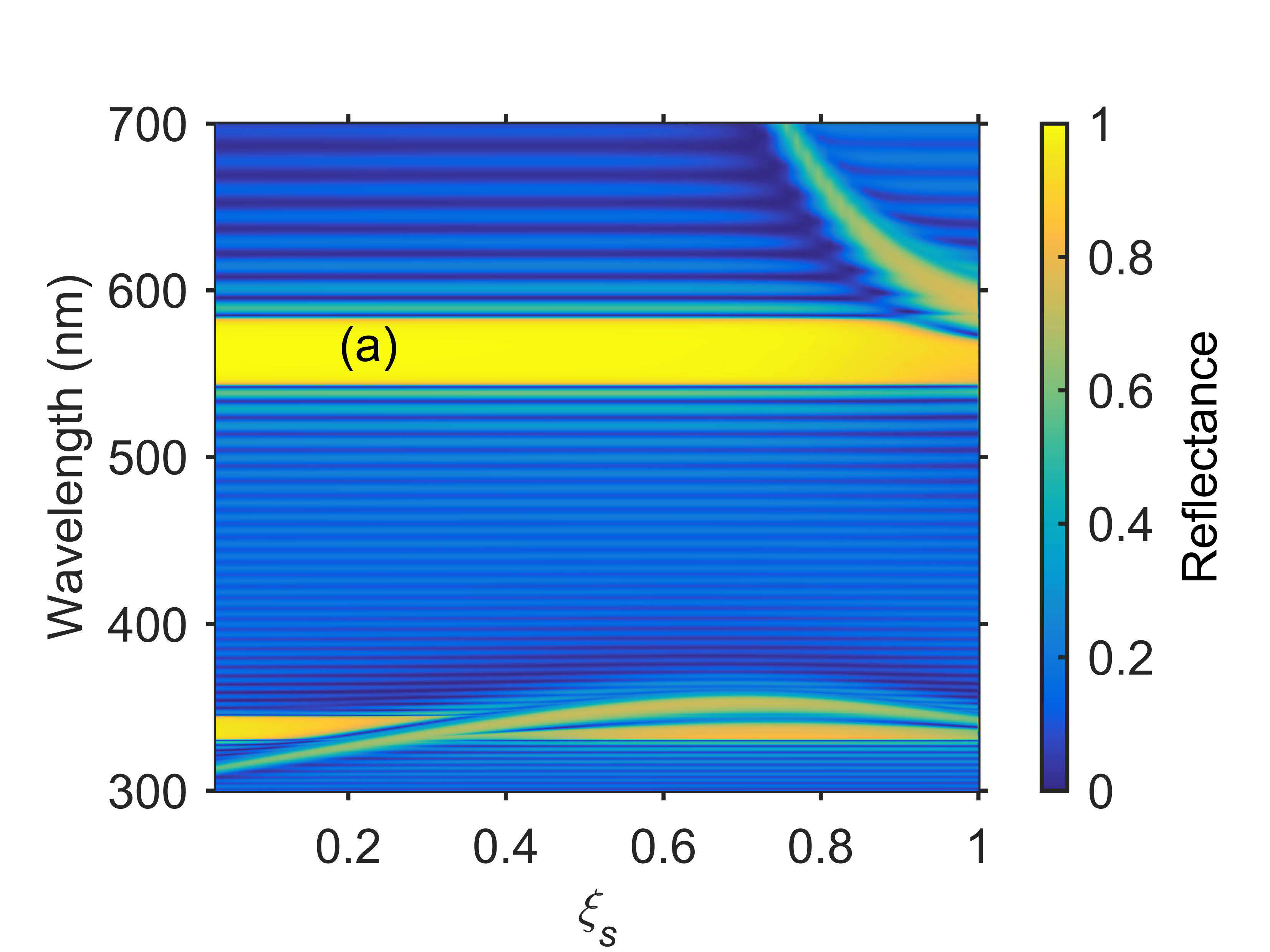}
\includegraphics[scale=1]{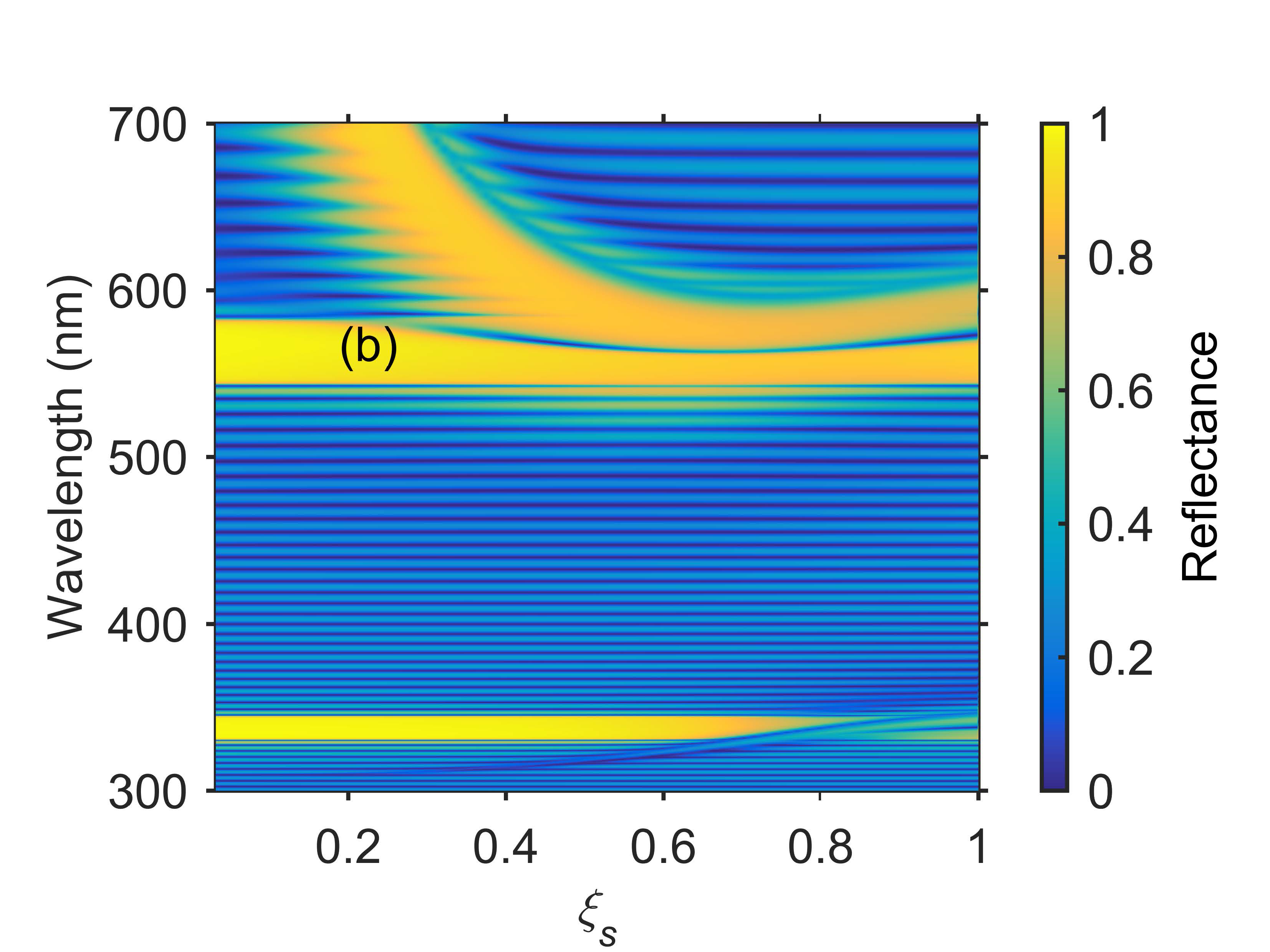}\\
Figure 3. Reflectance spectrum of the structure for (a) the $x$- and (b) $y$-polarized light as a function of the ratio $\xi_{s}$ between the particle axes at $\beta = 0.275$ and $f=0.08$. 
\label{fig:Spectra}
\end{center}

In the figures, one can see the static reflectance bands corresponding to the third (long-wavelength) and fifth (short-wavelength) PBGs with central wavelengths of $\lambda_3 = 4n_a d_a/3 = 563$ nm and $\lambda_5 = 4n_a d_a/5 = 338$ nm. In addition, there are reflectance bands corresponding to the NC layer. These bands repeat the behavior of the resonance NC permittivity regions in Fig. \ref{fig:Permittivity}. In the regions of overlap of the PC and NC reflection bands, resonance dips in the reflection are observed, which correspond to the localized modes in the structure. The NC permittivity anisotropy leads to the significant polarization sensitivity of the spectra and different behavior of the localized modes.

\begin{center}
\includegraphics[scale=1]{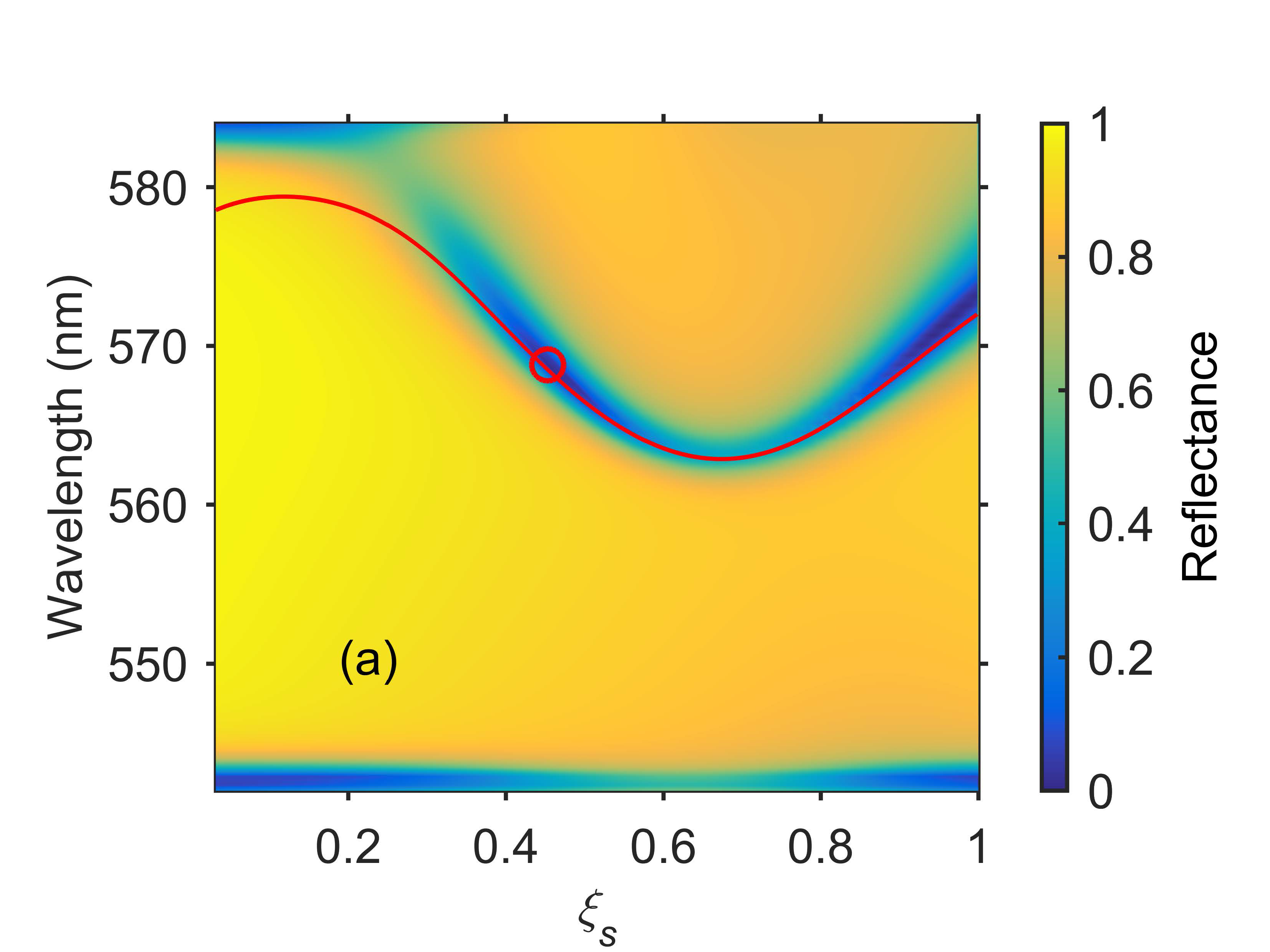}
\includegraphics[scale=1]{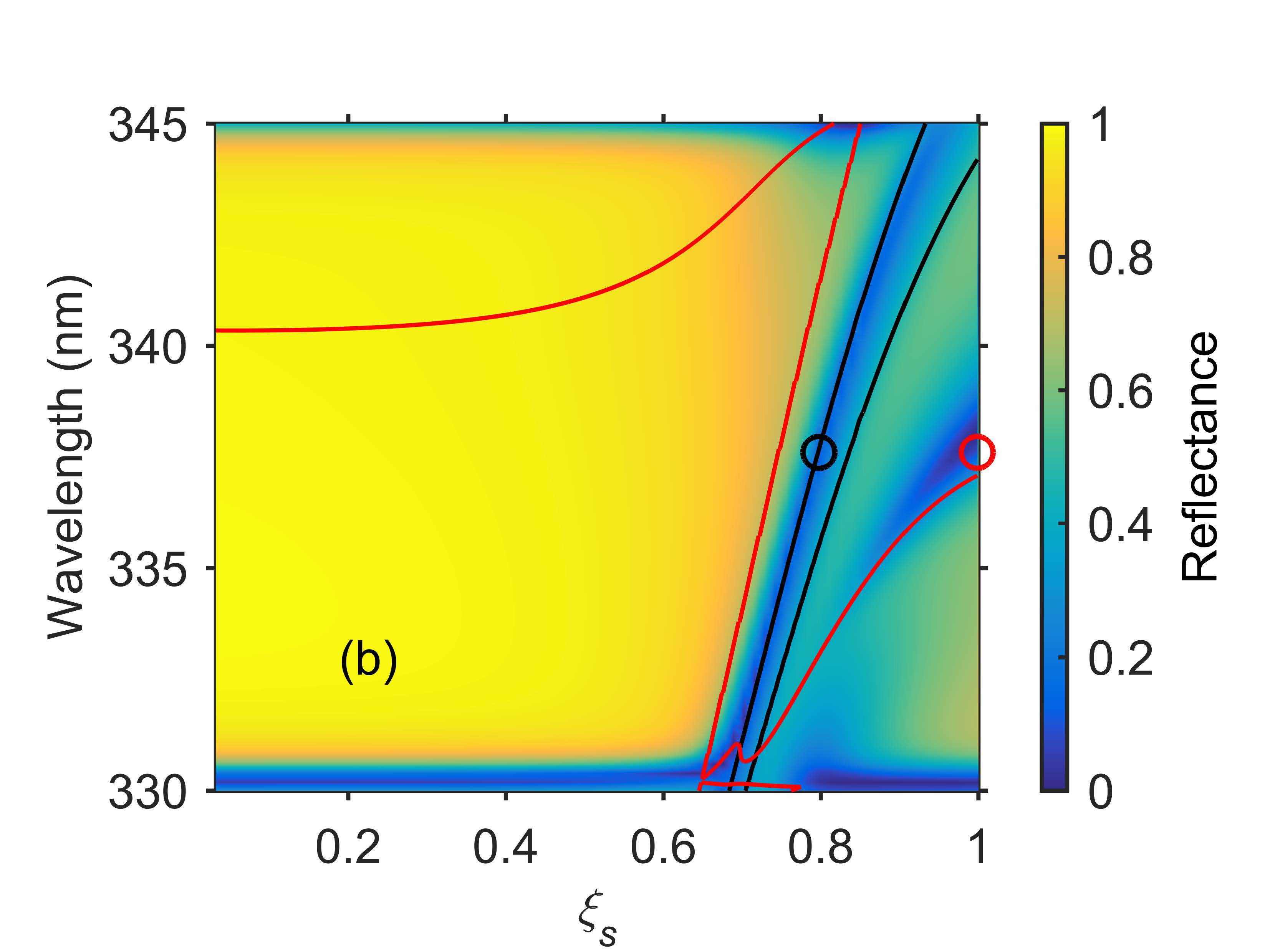}
\includegraphics[scale=1]{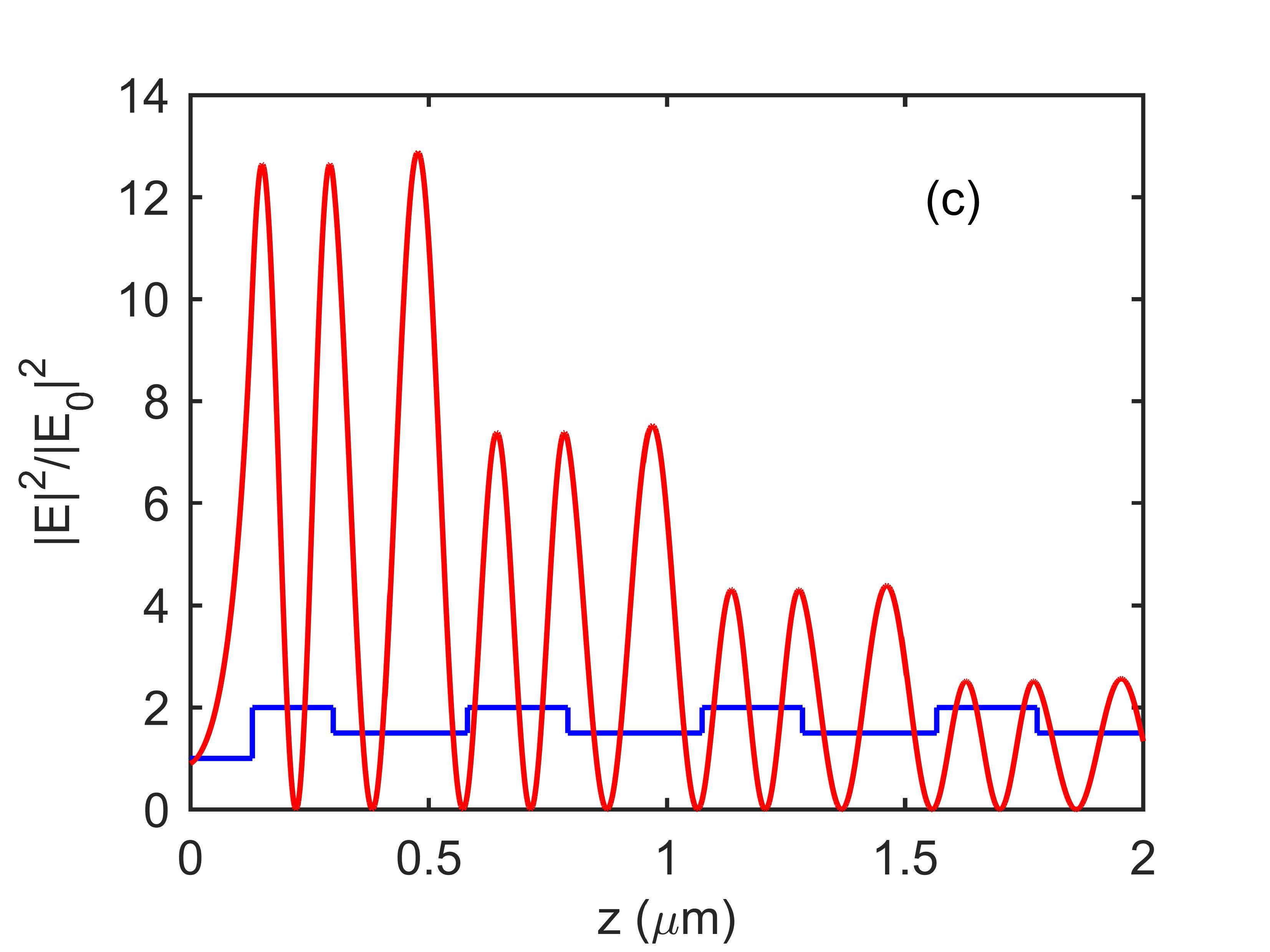}
\includegraphics[scale=1]{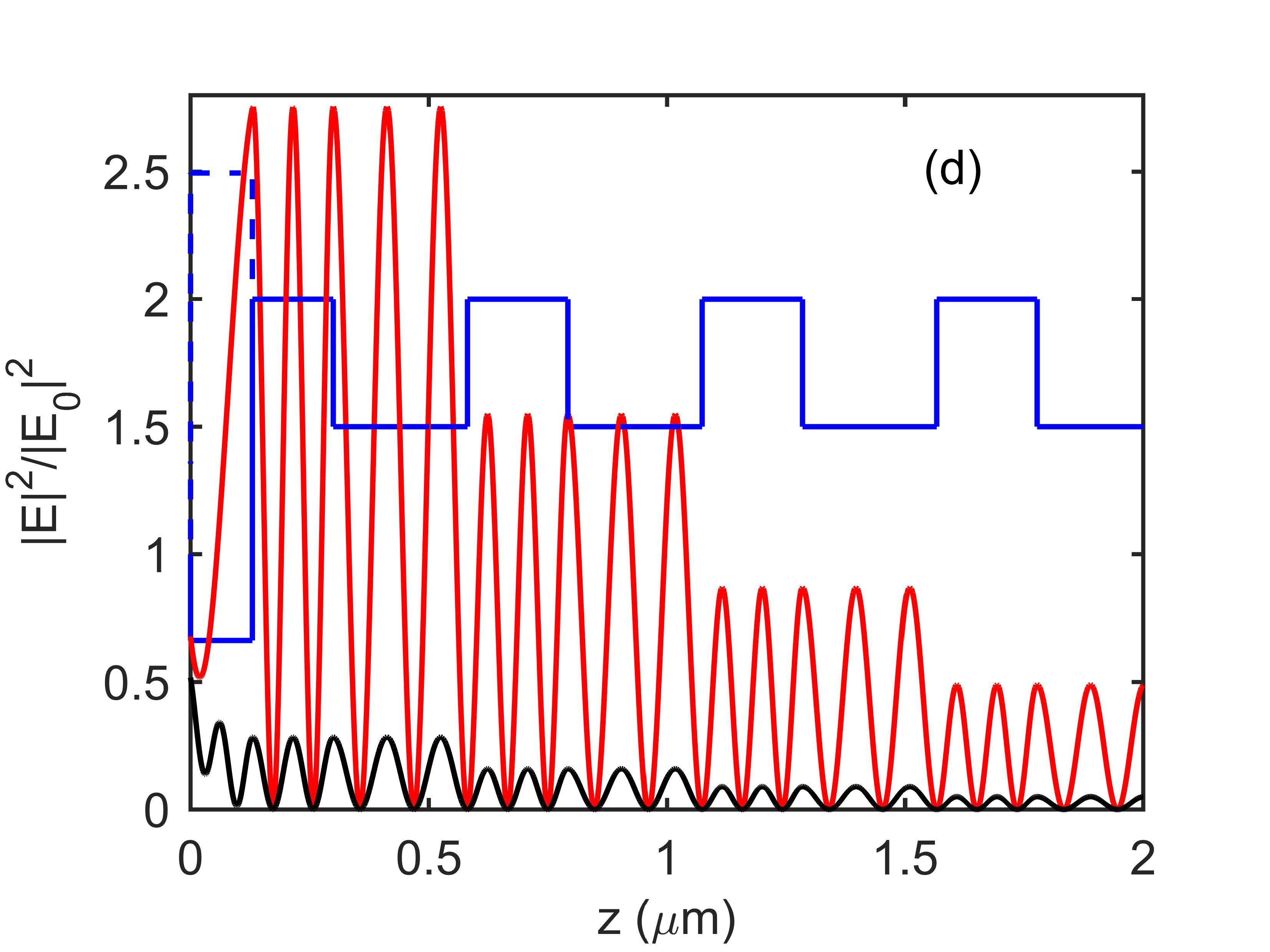}\\
Figure 4. (a, b) Reflectance spectrum of the structure for the $y$-polarized light as a function of the ratio $\xi_{s}$ between the particle axes at $\beta = 0.275$ and $f = 0.08$. The red line shows the solution of dispersion equation \ref{eq:TPP_disp} for the TPP and the black line, for the Fabry--Perot mode \ref{eq:FP_disp}. (c) Refractive index (blue) and local field intensity (red) distributions over the structure at parameters of (0.45, 568.8) (red circle in Fig. (a)). (d) Refractive index (blue) and local field intensity (red) distributions over the structure at parameters of (1, 337.6) (red circle in Fig. (b)) and  refractive index (blue dashed line) and local field intensity (black) distributions over the structure at parameters of (0.79, 337.6) (black circle in Fig. (b)).
\label{fig:Dispersion}
\item
\end{center}

Let us investigate the localized modes in more detail by the example of the $y$-polarized reflectance spectrum. The modes excited at $\xi_{s} = 1$ in both PBGs are the TPPs corresponding to the relatively high quality factor resonances. The TPP dispersion curve is obtained by solving the equation corresponding to the phase matching condition \cite{Kaliteevski2007}

\begin{equation}
\varphi_{NC} + \varphi_{PC} + \varphi_{Top} = 2\pi m,
\label{eq:TPP_disp}
\end{equation}
where $\varphi_{NC}$ is the phase shift upon reflection of light from the NC layer, $\varphi_{PC}$ is the phase shift upon reflection of light from the PC, $\varphi_{Top}$ is the phase incursion during back and forth light passage in the first PC layer adjacent to the NC layer, and $m$ is the integer. 
The first two terms in Eq. \ref{eq:TPP_disp} are determined as arguments of the complex amplitudes $r_{NC}$ and $r_{PC}$ of reflection from the NC and PC: $\varphi_{NC} = arg(r_{NC})$ and $\varphi_{PC} = arg(r_{PC})$. The amplitude $r_{NC}$ is expressed by Airy formula \cite{Born_Wolf1999b} for light falling onto the NC film with the permittivity $n_2 = \sqrt{\varepsilon_{yy}}$, which is placed between media with the permittivities $n_1 = n_a$ and $n_3 = n_{Air}$:

\begin{equation}
r_{NC}=\frac{r_{12}+r_{23}e^{2i\delta}}{1+r_{12}r_{23}e^{2i\delta}},\;
r_{12}=\frac{n_1-n_2}{n_1+n_2},\;
r_{23}=\frac{n_2-n_3}{n_2+n_3},\;
\delta=2 \pi n_2 d_{NC} \omega /c.
\label{eq:Airy}
\end{equation}
The amplitude $r_{PC}$ is determined from the well-known formula for a multilayer Bragg reflector  \cite{yariv1984optical}. The phase incursion in the first layer is $\varphi_{Top} = 4 \pi n_a d_{Top} \omega /c$.

Equation \ref{eq:TPP_disp} suggests that the first PC layer serving as a cavity should contain an integer number of half-wavelengths of the standing light wave. This equation determines the TPP wavelength; however, it is insufficient for the TPP formation. In addition, the NC and PC must have good reflection characteristics at the TPP wavelength; i. e., the amplitudes of reflection of light from them must maximally satisfy the condition $|r_{NC}||r_{PC}| \approx 1$ \cite{Kaliteevski2007}. It means that the TPP is excited in the region of overlap of the NC and PC reflection bands.   

Figures \ref{fig:Dispersion}a,b show the enlarged reflectance spectra for the $y$-polarized light in the PBG region of a PC. The red color corresponds to the solution of dispersion equation \ref{eq:TPP_disp} (at $m = 1$ in the long-wavelength PBG and $m = 2$ in the short-wavelength PBG), which agrees well with the resonance spectral manifestations of the TPP in both PBGs. The field distribution at the resonance wavelengths (Fig. \ref{fig:Dispersion}c,d) also corresponds to the TPP: the field is localized at the NC/PC interface and decreases exponentially deep in the NC and PC on either side from the interface. 

It is worth noting that the value of light localization at the TPP wavelength in the long-wavelength PBG (Fig. \ref{fig:Dispersion}c) is larger than in the short-wavelength PBG by an order of magnitude (Fig. \ref{fig:Dispersion}d). This is related to the difference between the amplitudes of reflection from the NC: at a wavelength of $\lambda = 568.8$ nm corresponding to the long-wavelength TPP (Fig. \ref{fig:Dispersion}c), we have $|r_{NC}| = 0.85$ (at $\xi_{s} = 0.45$) and, at a wavelength of $\lambda = 337.6$ nm corresponding to the short-wavelength TPP (Fig. \ref{fig:Dispersion}d), $|r_{NC}| = 0.6$ (at $\xi_{s} = 1$). In addition, we would like to mention the peculiarities of reflection of light from the NC for both TPP modes. For the long-wavelength TPP, the NC is similar to a metal and has a permittivity of $\varepsilon_{yy} = -0.9860 + 0.1973i$ at a wavelength of $\lambda = 568.8$ nm, while for the short-wavelength TPP, the NC is similar to a dissipative dielectric \cite{Bikbaev2017} and has a permittivity of $\varepsilon_{yy} = 0.3502 + 0.2659i$ at a wavelength of $\lambda =337.6$ nm.    

The mode excited at $0.7 < \xi_{s} < 0.9$ in the short-wavelength PBG is the Fabry--Perot cavity eigenmode, which corresponds to the resonance with a relatively low quality factor. The Fabry--Perot cavity is formed by an NC layer localized between air and the first PC layer \cite{Bikbaev2018}. The dispersion curve of the Fabry--Perot mode is determined by solving the equation corresponding to the phase matching condition

\begin{equation}
-r_{12} + r_{23} + 2\delta = 2\pi m.
\label{eq:FP_disp}
\end{equation}
Figure \ref{fig:Dispersion}b illustrates the solution of dispersion equation \ref{eq:FP_disp} at $m = 2$ (shown with black). One can see good agreement between the resonance spectral manifestation of the Fabry--Perot mode. The field distribution at the wavelength of this resonance (Fig. \ref{fig:Dispersion}d) also corresponds to the Fabry--Perot mode. It can be seen that, in this case, the NC layer contains an integer number of the incident radiation wavelengths. The field localization in the NC is relatively weak because of the poor reflection properties of the NC layer boundary: the reflectivities are $|r_{12}| = 0.13$ and $|r_{23}| = 0.43$  at a wavelength of  $\lambda = 337.6$ nm (at $\xi_{s} = 0.79$). Similarly, we can analyze the $x$-polarized spectrum of the structure (Fig. \ref{fig:Spectra}a). In this case, the TPP also manifests itself in the spectrum in the long-wavelength region, while in the short-wavelength region both the TPP and the Fabry--Perot mode are observed.

\section{Conclusions}

The spectral manifestation of the Tamm plasmon polaritons localized at the interface between the nanocomposite and photonic crystal was numerically studied.
The nanocomposite consists of particles formed from a dielectric core and a silver shell uniformly dispersed in a transparent matrix.

The occurrence of the polarization sensitivity of the spectra with the change in the nanoparticle shape from spherical to spheroidal was demonstrated.
The change in the spectra with an increase in the degree of flattening of the nanospheroids was examined.

It was shown that, along with the Tamm plasmon in the photonic band gap, there is also a Fabry--Perot mode, the light field of which is localized inside the nanocomposite layer.  

The dispersion curves presented for the Tamm plasmon and Fabry--Perot mode agree well with the spectral manifestations of the modes obtained by the transfer matrix method.

The demonstrated high polarization sensitivity is expected to be used in Tamm-plasmon-based devices.

\ack

\hspace*{\parindent} 
This study was supported by the Russian Foundation for Basic Research, the Government of the Krasnoyarsk Territory, and the Krasnoyarsk Territorial Foundation for Support of Scientific and R\&D Activities, project no.\textsuperscript{\underline{o}} 17-42-240464. P. S. P. acknowledges the support of the Scholarship of the President of the Russian Federation no.\textsuperscript{\underline{o}} SP-227.2016.5.

\section*{References}

\input{CoreShellSpher.bbl}

\end{document}

%% file: CoreShellSpher.bbl
\providecommand{\newblock}{}